\documentclass[aip,amsmath,amssymb,graphicx,reprint]{revtex4-1}

\usepackage{graphicx}
\usepackage{dcolumn}
\usepackage{bm}
\usepackage{caption}
\usepackage{subcaption}
\usepackage{xcolor}
\usepackage{physics}
\usepackage{mathtools}
\usepackage{ulem}
\usepackage{placeins}

\begin{document}


\title{Quasi-Helmholtz Decomposition, Gauss' Laws and Charge Conservation for Finite Element Particle-in-Cell} 



\author{ Scott O'Connor}
\altaffiliation[Also at ]{Department of Computational Science, Mathematics, and Engineering, Michigan State University, East Lansing, MI}
\author{Zane D. Crawford}
\altaffiliation[Also at ]{Department of Computational Science, Mathematics, and Engineering, Michigan State University, East Lansing, MI}
\author{O. H. Ramachandran}
\altaffiliation[Also at ]{Department of Computational Science, Mathematics, and Engineering, Michigan State University, East Lansing, MI}
\author{John Luginsland}
\affiliation{Department of Electrical and Computer Engineering, Michigan State University, East Lansing, MI}
\author{ B. Shanker}
\affiliation{Department of Electrical and Computer Engineering, Michigan State University, East Lansing, MI}

\date{\today}

\begin{abstract}

Development of particle in cell methods using finite element based methods (FEMs) have been a topic of renewed interest; this has largely been driven by (a) the ability of finite element methods to better model geometry, (b) better understanding of function spaces that are necessary to represent \emph{all} Maxwell quantities, and (c) more recently, the fundamental rubrics that should be obeyed in space and time so as to satisfy Gauss' laws and the equation of continuity.  In that vein, methods have been developed recently that satisfy these equations and are \emph{agnostic} to time stepping methods. While is development is indeed a significant advance, it should be noted that implicit FEM transient solvers support an underlying null space that corresponds to a gradient of a scalar potential $\grad \Phi(\vb{r})$  (or $t \grad \Phi (\vb{r})$ in the case of wave equation solvers). While explicit schemes do not suffer from this drawback, they are only conditionally stable, time step sizes are mesh dependent, and very small.  A way to overcome this bottleneck, and indeed, satisfy \emph{all} four Maxwell's equation is to use a quasi-Helmholtz formulation on a tesselation. In the re-formulation presented, we strictly satisfy the equation of continuity and Gauss' laws for both the electric and magnetic flux densities. Results demonstrating the efficacy of this scheme will be presented. 
\end{abstract}


\maketitle 

\section{Introduction}

Simulation of plasma and space charge has many applications in science and technology ranging from particle accelerators to satellites and medicine\cite{marchand2011ptetra,lemke1999three,fourkal2002particle}.  A popular method for simulation of plasma is particle-in-cell (PIC) which self-consistently evolves particle motion for charge species with Maxwell’s Equations. 
Traditionally PIC has been based on finite difference time domain with the domain represented by Yee cells\cite{yee1966numerical}. The stair stepping nature of a structured grid presents challenges on geometry modeling especially fine features and multi-scale objects. Methods such as cut-cells\cite{grote2005warp} have been developed to alleviate this challenge. Another fundamental bottleneck is that these methods are conditionally stable, with time step size of the field solver dictated by the mesh, and is relatively small so as to satisfy the Courant-Friedrich-Levy condition. As a result, there has been renewed interest in developing methods that not only better represent the geometry but better capture the underlying physics. 

The route taken is to use finite element methods (FEMS); they have an extensive history of use in field solvers due to their ability to better represent the geometry and physics. But more importantly, there is a well developed rigorous body of literature \cite{monk2003finite,bossavit1991student,bossavitwhitney}  on the mathematical foundation of FEM.  Indeed, as is to be expected, FEM has made inroads into PIC modeling more of late with some excellent work rubrics of this approach\cite{perse2021geometric,ricketson2020energy,he2016hamiltonian,hirvijoki2020subcycling,evstatiev2013variational,xiao2016explicit,pinto2021variational,morrison2017structure,jianyuan2018structure,kormann2021energy,burby2017finite,qin2015canonical,he2015hamiltonian,shadwick2014variational,crouseilles2015hamiltonian,stamm2014variational,xiao2013variational}. Of particular note in seminal work in \cite{pinto2014charge}, wherein rubrics for charge conservation for FEM-PIC where developed. These have been refined and developed further in a series \cite{moon2015exact,na2017axisymmetric,na2018relativistic}. By and large, these papers follow a similar rubric, use a leap-frog scheme to step through a solution for the field, update position and velocity of the particle and solve for the fields again. In this setting, it was shown that the methodology is charge conserving. While this the classical PIC cycle, a more recent trend has been to solve both the electromagnetic and Newton's equations self-consistently at the same time step\cite{kormann2021energy}. This approach necessary when the motion is close to being relativistic. 

In the problems that we address in this paper, our motion is non-relativistic. As a result, we are in the regime where the classic PIC is valid. Even so, being restricted to a time step size that is tightly related to the mesh is undesirable, especially, when there are a number of well established methods that make the field solver unconditionally stable \cite{zienkiewicz1977new}. Unfortunately, developing a method where Gauss' laws and equation of continuity is satisfied is not a trivial task. As an aside, in the rest of the paper, we will use the term \emph{conservation laws} to denote these equations and admit to an abuse of terminology! This problem was solved relatively recently \cite{2021arXiv210206248O,crawford2021rubrics} with results demonstrating satisfaction of conservation laws to almost machine precision. But this is not really the end of the story. While leap frog FEM is known to not have a null space, implicit time stepping methods do. There are two flavors of time domin FEM-those that discretize Maxwell's equations directly (MFEM) and those that solve the wave equation (WE-FEM). These equation suffer from a null space that behave as a gradient of a potential and a product time with a gradient of a potential, respectively. As a result, it is possible to add spurious charge into the system corresponding to the divergence of these null-spaces. This is especially true when one uses iterative methods to solve the system of equations where level of excitation of the null space depends on the error threshold. Given that satisfaction conservation laws is paramount in a PIC solution, we prescribe an alternate solution. 

Traditional solution relies on solving for fields via the two curl equations. Gauss' laws will be satisfied by these fields provided equations of continuity hold and there are no null spaces excited. But this is easier said than done. It follows that perhaps an alternative is to use a Helmholtz decomposition of fields in the construction of field solvers. This would permit solution to all four equations. Means to effect this is the overall thrust of this paper. 

It is well known that an exact Helmholtz decomposition is not possible in a discrete setting. Indeed, using an analytical decomposition as a starting point \emph{prior} to discretization will lead to erroneous results as one needs to ensure that proper mapping on \emph{de-Rham} spaces are satisfied. What one can achieve is a quasi-Helmholtz decomposition where one of the two conditions are strongly satisfied. The development of quasi-Helmholtz decomposition dates back to the early days of development  vector basis for modeling electromagnetic fields \cite{kotiuga1984hodge,bossavitwhitney,kettunen1998discrete,bossavit1998geometry}. 
Seminal literature was devoted to understanding the necessary mathematics in the language of differential forms and was driven by the need to solve magnetostatic problems. Development of quasi-Helmholtz decomposition can be traced back to tree-cotree decomposition for magnetostatics \cite{andriulli2012loop,andriulli2012well,manges1995generalized,manges1997tree,wang2010application}. Development of of this body of work has progressed from simply connected to multiply-connected objects\cite{suuriniemi2002generalization,kettunen1998formulation}. As an aside, this period was rich with development of state of the art FEM techniques, including MFEM leap frog methods \cite{wong1995finite}.

This body of work is the starting point of the work presented in this paper. We will restrict ourselves to simply connected objects and zero-th order basis function; extension to multiply connected objects and higher order basis is underway and will be presented elsewhere. In this paper, our principal contributions of this paper are as follows: 
\begin{enumerate}
    \item We will develop quasi-Helmholtz projectors to partition fields into solenoidal and non-solenoidal components for both the electric field and magnetic flux density. 
    \item In specializing these to simply connected systems, we will use a co-tree identification of degrees of freedom such that the total number of degrees of freedom is identical to the original system. 
    \item We will show that the resulting system of equations map exactly to Gauss' law and the equation of continuity is exactly satisfied. 
    \item We apply these to both MFEM and WE-FEM, discuss ramification of null spaces and show connections/variations from the well known $\vb{A}-\Phi$ formulation.  
    \item Finally, we will present numerous results that validate the arguments presented in the paper. 
\end{enumerate}

The rest of this paper is organized as follows: Next, in Section \ref{sec:formulation} we present the desired formulation of the PIC system in the continuous domain. Section \ref{sec:quasi-helmholtz} presents the qausi-Helmholtz decomposition in a discrete setting for both the electric field and magnetic flux density, and derivation of the revised sets of equation. We also show how these equations satisfy conservation laws. Next, in Section \ref{sec:nullSpaces}, we discuss null spaces that arise out of these equations as well as gauge considerations. Finally, results are presented in in \ref{sec:results} that demonstrates the advantages of the proposed method.

\section{Formulation\label{sec:formulation}}
\subsection{Preliminaries}

Consider a domain $\Omega$ whose boundaries are denoted by $\partial \Omega$ and enclose the domain.
The domain consists of free space and permittivity $\varepsilon_0$ and permeability $\mu_0$ with the speed of light denoted using $c = 1/\sqrt{\mu_0\epsilon_0}$. 
The enclosing boundaries are either Nuemann or Dirichlet. 
The excitation of the system comes from electromagnetic field due to moving charges whose position (and velocity) evolve over.
The moving charges are assumed to be collisionless and follows the Vlasov equation,
\begin{align} \label{eq:vlasov}
  \partial_t f(t,\vb{r},\vb{v})  + \vb{v} \cdot \nabla f(t,\vb{r},\vb{v}) + \\ \frac{q}{m} [\vb{E}(t,\vb{r}) + \vb{v} \times \vb{B}(t,\vb{r})] \cdot \nabla_v f(t,\vb{r},\vb{v}) = 0, \nonumber
\end{align}
where $f(t,\vb{r},\vb{v})$ is a phase space distribution function (PSDF).
 We follow the conventional definition of the charge and current density defined as $\rho(t,\vb{r})=q\int_{\Omega} f(t,\vb{r},\vb{v})d\vb{v}$ and $\vb{J}(t,\vb{r})=q\int _{\Omega}\vb{v}(t)f(t,\vb{r},\vb{v})d\vb{v}$ as moments of the PSDF \cite{bittencourt2013fundamentals}.
 The fields, $\vb{E}(t,\vb{r})$ and $\vb{B}(t,\vb{r})$, in Vaslov equation are solutions to the Maxwell's equations Eq. \ref{eq:maxwell}.
\begin{subequations}\label{eq:maxwell}
  \begin{align}
        \div \vb{D}(t,\vb{r}) & = \rho (t,\vb{r}) \label{eq:Gauss_Law} \\ 
        \div \vb{B}(t,\vb{r}) & = 0 \label{eq:Gauss_magnetic_law}\\ 
       \curl \vb{E}(t,\vb{r}) & = - \partial_t \vb{B}(t,\vb{r}) \label{eq:Faradays_Law}\\ 
        \curl \vb{H}(t,\vb{r})& = \partial_t  \vb{D}(t,\vb{r}) + \vb{J}(t,\vb{r}) \label{eq:Ampheres_Law}
    \end{align}
\end{subequations}
where $\vb{D}(t,\vb{r}) = \varepsilon\vb{E}(t,\vb{r})$ and $\vb{H}(t,\vb{r}) = \mu^{-1}\vb{B}(t,\vb{r})$.
The boundary conditions enclosing the domain are either Neumann $\partial \Omega_N$ or Dirichlet $\partial \Omega_D$,
\begin{subequations}\label{eq:bceq}
\begin{align}
\hat{n}\times \mathbf{E}(t, \mathbf{r}) &= \mathbf{\Psi}_D(t, \mathbf{r})\;\;\text{on}\;\partial\Omega_D,\label{eq:dirichlet} \\
\hat{n}\times \mu^{-1}\mathbf{B}(t, \mathbf{r}) &= \mathbf{\Psi}_N(t, \mathbf{r})\;\;\text{on}\;\partial\Omega_N\label{eq:neumann},
\end{align}
\end{subequations}
 where $\hat{n}$ is an outward pointing normal to $\partial \Omega_N$ and $\partial \Omega_D$ and $\mathbf{\Psi}_N(t, \mathbf{r})$ and $\mathbf{\Psi}_D(t, \mathbf{r})$ are Neumann and Dirichlet boundary condition functions.
In addition, it can be shown from Maxwell's equations that the equation of continuity
\begin{align}\label{eq:continuity}
    \nabla \cdot \vb{J}(t, \mathbf{r}) + \partial_t \rho(t, \mathbf{r}) = 0.
\end{align}
  holds. While $f(t,\vb{r},\vb{v})$ is not solved directly, as it is computationally prohibitive, we use an approach where one approximates the PSDF as a collection of discrete point particles. 
We follow the usual representation where  $\rho(t,\vb{r}) = q\sum^{N_p}_{p=1} \delta(\vb{r}-\vb{r}_p (t))$ and $\vb{J}(t,\vb{r}) = q\sum^{N_p}_{p=1}  \vb{v}_p(t) \delta(\vb{r}-\vb{r}_p (t) )$.  Generalization to other shape function (instead of delta functions) is certainly possible\cite{crawford2021rubrics}, but is beyond the scope of this paper and not pertinent to the central thesis of this paper. As in PIC schemes, the particles  are moved using the Lorentz force $\vb{F} (t, \vb{r})  = q(\vb{E}(t, \vb{r}) + \vb{v}(t, \vb{r})\times \vb{B}(t, \vb{r}))$ and Newton's equations. A self-consistent solution to both evolution of charges and fields constitute a PIC methodology.

\subsection{Helmholtz Decomposition}\label{sec:helmholtz}

It is well established via Helmholtz theorem that any sufficiently smooth vector field $\bm{\Lambda} (t, \mathbf{r}) = \bm{\Lambda}_{ir} (t, \mathbf{r}) + \bm{\Lambda}_{r} (t, \mathbf{r}) $ can be represented into a rotational (divergence-free) $\bm{\Lambda}_{r}(t, \mathbf{r})$ and irrotational (curl-free) $\bm{\Lambda}_{ir}(t, \mathbf{r})$ components. This implies that $\nabla \times \bm{\Lambda}_{ir}(t, \mathbf{r})=0$ and $\nabla \cdot \bm{\Lambda}_{r}(t, \mathbf{r})=0$.  Along this vein, it follows that both the electric and magnetic fields can be decomposed in a similar manner. Parenthetically, we note here that in the PIC context, our point charges are measured on the mesh, thereby smoothing the sources and the resulting vector field. Given this decomposition we can rewrite Maxwell's equations as, 
\begin{subequations}
    \begin{align}
        \div\vb{D}_{ir} (t, \vb{r})  & = \rho (t, \vb{r})\\
        \div \vb{B}_{r} (t, \vb{r}) & = 0 \\
        \curl \vb{E}_{r} (t, \vb{r}) & =  - \partial_t \vb{B}_{r} (t, \vb{r}) \\
        \curl \vb{H}_{r} (t, \vb{r}) & =    \partial_t\vb{D}_{r} (t, \vb{r})+\partial_t\vb{D}_{ir}(t, \vb{r})  + \vb{J} (t, \vb{r})
    \end{align}
\end{subequations}
Additionally the continuity equation is updated as, $\nabla \cdot \vb{J}_{ir} (t, \vb{r})+ \partial_t \rho (t ,\vb{r}) = 0$
The main takeaways are the following; (a)  the three components of fields that one needs to determine are $\vb{E}_{ir} (t,\vb{r})$, $\vb{E}_{r}(t,\vb{r})$ and $\vb{B}_{r}(t,\vb{r})$,  (b) Maxwell's equations are sufficient to determining these components, and (c) solving these would ensure that null-spaces do not corrupt conservation laws. Of course, it is well known that it is not possible to construct a complete Helmholz decomposition in a discrete setting. As result, in what follows, we will develop a quasi-Helmholtz decomposition on discretization of the domain. For simplicity of exposition, we  assume that the domain is simply connected \cite{bossavitwhitney} . 

\subsection{Discrete Framework}

The analysis framework starts with the \emph{de-Rham} complex\cite{bossavitwhitney} to represent fields and fluxes. To that end, we assume that the domain is represented using a collection of finite elements $\mathcal{K} = \left \{ \mathcal{N}, \mathcal{E}, \mathcal{F}, \mathcal{T} \right \}$ defined using $N_n$ nodes, $N_e$ edges, $N_f$ faces and $N_t$ tetrahedron. 
On these elements one can define Whitney spaces \cite{bossavitwhitney,monk2003finite} $W^0(\vb{r})$, $\vb{W}^1(\vb{r})$, $\vb{W}^2(\vb{r})$ and $W^3(\vb{r})$ to represent quantities on the primal grid. In what follows, we represent electric fields using  $\vb{E} (t, \vb{r}) = span \left \{ \vb{W}^1 (\vb{r})\right \} $ and $\vb{B} (t, \vb{r}) = span \left \{ \vb{W}^2 (\vb{r}) \right \} $.  Specifically,    $\vb{E}(t,\vb{r}) = \sum_{i=1}^{N_e} e_i(t) \vb{W}^{(1)}_{i}(\vb{r})$ and   $\vb{B}(t,\vb{r}) = \sum_{i=1}^{N_f} b_i(t) \vb{W}^{(2)}_{i}(\vb{r})$. It follows that these coefficients can be arranged into vector of the form
 $\bar{E}(t) = [e_1(t),e_2(t),\dots,e_{N_e}(t)]$, $\bar{B}(t) = [b_1(t), b_2(t),\dots,b_{N_f}(t)]$; the value at any time $n \Delta_t$ is denoted using $\bar{E}(n\Delta_t) = \bar{E}^n$ (likewise for $\bar{B}(n\Delta_t) = \bar{B}^n$) where $\Delta_t$ is the time step size. Similar notation is adopted for the coefficients of both the magnetic field and electric flux density. 
 
 In keeping with this framework, it follows that the electric flux densities and magnetic field lies in the dual space, but can be presented using primal space quantities. But before we do so, we define the following Hodge matrices $[\star_{\epsilon}]_{i,j}$ and $[\star_{{\mu}^{-1}}]$ that map  primal grid quanties to the dual grid:  
\begin{align}
        [\star_\epsilon]_{i,j} &= \langle \vb{W}^{(1)}_i(\vb{r}),\varepsilon\cdot\vb{W}^{(1)}_j(\vb{r}) \rangle;  i,j \in \mathcal{E} \\
        [\star_{\mu^{-1}}]_{i,j} &= \langle \vb{W}^{(2)}_i(\vb{r}),\mu^{-1}\cdot\vb{W}^{(2)}_j(\vb{r})\rangle; i,j \in \mathcal{F}\\
        [\star_{\rho}]_{i,j} &= \langle W^{(3)}_i(\vb{r}),W^{(3)}_j(\vb{r})\rangle; i,j \in \mathcal{T}
\end{align}
In addition, we define the following
\begin{subequations}
\begin{align}
        [\vb{M}_g]_{i,j} &= \langle \vb{W}^{(1)}_i (\vb{r}) , \nabla W^{(0)}_j (\vb{r})\rangle; i \in \mathcal{E}, j \in \mathcal{N} \label{eq:mg}\\
        [\vb{M}_c]_{i,j} &= \langle  \vb{W}^{(2)}_i (\vb{r}), \curl \vb{W}^{(1)}_j (\vb{r})\rangle;  i \in \mathcal{F}, j \in \mathcal{E} \label{eq:mc}\\
        [\vb{M}_d]_{i,j} &= \langle W^{(3)}_i(\vb{r}) , \div \vb{W}^{(2)}_j (\vb{r}) \rangle;  i \in \mathcal{T}, j \in \mathcal{F} \label{eq:mc}\\
        [\grad ] &= \varepsilon[\star_{\epsilon}]^{-1}[\vb{M}_g] \label{eq:grad_mat}\\
        [\curl ] &= \mu^{-1}[\star_{\mu^{-1}}]^{-1}[\vb{M}_c]\label{eq:curl_mat}\\
        [\div ] & = [\star_{\rho}]^{-1}[\vb{M}_d] \label{eq:div_mat}
\end{align}
\end{subequations}
where $[\curl ]$ is the discrete curl operator, $[\grad ]$ is the discrete gradient operator, and $[\div]$ is the discrete divergence operator. We note that using this notation, one can obtain the coefficients for the electric flux density as 
$\bar{D}^n  = [\star_\epsilon] \bar{E}^n$. 
Using this framework, one can write the discretized equations in space as 
\begin{align}\label{eq:maxwell_semi}
\mqty[ [I]& 0 \\ 0 &[\star_{\varepsilon}]] \mqty[\partial_t \bar{B}\\ \partial_t \bar{E} ] + \mqty[0&  [\curl] \\ -[\curl]^T [\star_{\mu^{-1}}]& 0 ] \mqty[\bar{B} \\\bar{E} ] = \mqty[0\\ -\bar{J} ]
\end{align}
where $\bar{J}=[j_1(t),j_2(t),...j_{N_e}(t)]$ with $j_i (t) = \langle \vb{W}_i^{(1)} (\vb{r}), \vb{J} (t, \vb{r}) \rangle $, and $\bar{J}^n = \bar{J}(n\Delta_t)$.
Likewise, one can arrive at a similar discrete system for the wave equation as
\begin{align}\label{eq:wave_cont}
[\star_{\varepsilon}] \partial_t^2 \bar{E}  + [\star_S]\bar{E} = -\partial_t\bar{J} 
\end{align}
where $[\star_S]=[\curl]^T[\star_{\mu^{-1}}][\curl]$.
Using these equations as a backdrop, we next discuss the steps  necessary to effect  a discrete quasi-Helmholtz decomposition. 

\section{Quasi-Helmholtz Decomposition}\label{sec:quasi-helmholtz}

As noted earlier, it is well known that one cannot develop a complete Helmholtz decomposition of fields. It follows, that we may be able to achieve one condition, viz., either divergence \emph{or} curl free fields. To this end, prior to developing a numerical framework, we ask what is necessary for the fields to satisfy. In plasma physics, it follows that one must satisfy the condition $\div \vb{D}_{r} (t,\vb{r}) = 0$ and $\vb{B}_{ns}(t,\vb{r}) = 0$ strongly; in other words, we cannot necessarily ensure $\curl \vb{D}_{ir} (t,\vb{r}) = 0$. As eloquently pointed out by Vecci \cite{vecchi1999loop}, the condition that we are strongly imposing is the solenoidal nature of $\vb{D}_r (t,\vb{r})$; i.e., $\vb{D}_{ir}(t,\vb{r})$ is non-solenoidal but not irrorational, i.e., $\curl \vb{D}_{ir}(t,\vb{r}) \neq 0$. Henceforth, we will use subscripts ``$s$'' and ``$ns$'' to denote solenoidal and non-solenoidal components, respectively.  

To discuss this decomposition, we start with the electric field. The usual starting point of the analysis is \emph{de-Rham} diagrams. We focus on the sequence  $W^0 \overset{\nabla}{\longrightarrow}\vb{W}^1$. It is apparent that in the framework being pursued we seek to exploit the representation of the potential $\phi (t,\vb{r}) = span \left \{ W^0 (\vb{r}) \right \}$. As $\grad W^0 \in \vb{W}^1$, and we seek a decomposition of the form $\vb{E}(t, \vb{r}) = \vb{E}_{ns} (t, \vb{r}) + \vb{E}_{s} (t, \vb{r}) $ it follows that these statements can be made in the discrete setting via
\begin{subequations}\label{eq:efld_decomp}
\begin{align}
\sum_{i=1}^{N_e} e_i^n \vb{W}_i^1 (\vb{r}) & = \grad \sum_{j = 1}^{N_n}e^n_{j,ns} W_j^0 (\vb{r})  + \vb{E}_{s} (n\Delta_t, \vb{r}) \label{eq:efld_decomp_full}\\
[\star_\varepsilon] \bar{E}^n & =  \varepsilon [\vb{M}_g]\bar{E}_{ns}^n + \Lambda \bar{E}_{s}^n  \label{eq:efld_decomp_matrix} \\
\bar{D}^n & =  \Sigma \bar{E}_{ns}^n + \Lambda \bar{E}_{s}^n \label{eq:efld_decomp_flux}
\end{align}
\end{subequations}
We have chosen not to represent $\vb{E}_{s} (n\Delta_t, \vb{r})$ explicitly, leaving it as the remainder. It follows, that one can choose basis sets (the so called co-tree basis), but that is not quite important right now. It is sufficient to say that there exists a matrix $\Lambda$ that is an inner product of these basis with  $\vb{W}^1 (\vb{r})$. 

In a similar manner, we seek a representation of the magnetic flux density such that $\vb{B} (t, \vb{r}) = \vb{B}_{ns} (t,\vb{r}) + \vb{B}_{s} (t,\vb{r})$. To do so we take recourse again to the \emph{de-Rham} picture and note that the magnetic field, $\vb{H} (t, \vb{r})$, lies the dual space $\widetilde{\vb{W}}^1 (\vb{r}) $, where \emph{tilde} denotes quantities associated with the dual space. As before, we take recourse to the sequence $\widetilde{W}^0 \overset{\nabla}{\longrightarrow}\widetilde{\vb{W}}^1$ and note that  $\widetilde{\grad}\widetilde{W}^0 (\vb{r}) \in \widetilde{\vb{W}}^1$.  In spirit of \eqref{eq:efld_decomp}, it follows that one can write
\begin{subequations}\label{eq:bfld_decomp}
\begin{align}
    \sum_{i=1}^{N_f} h_i^n \widetilde{\vb{W}}_i^1 (\vb{r}) & = \widetilde{\grad} \sum_{j = 1}^{N_t}h^n_{j,ns} \widetilde{W}_j^0 (\vb{r}) + \vb{H}_{s} (n\Delta_t, \vb{r}) \label{eq:bfld_decomp_full}\\
     [\tilde{\star}_{\mu}] \bar{H}^n & = \mu[\div]^T\bar{H}_{ns}^n + \mu\Lambda^\prime_m \bar{H}_{s}^n  \label{eq:efld_decomp_matrix}\\
    \bar{B}^n & =  \Sigma_m \bar{B}_{ns}^n + \Lambda_m \bar{B}_{s}^n \label{eq:efld_decomp_flux}
    \end{align}
\end{subequations}
where $[\tilde{\star}_{\mu}]\bar{H}^n = \bar{B}^n$. In all both \eqref{eq:efld_decomp} and \eqref{eq:bfld_decomp}, we require $\Sigma^T \Lambda = 0$ and $\Sigma_m^T \Lambda_m = 0$. This imposes the necessary solenoidal nature of the fields. Parenthetically, we note that $\Sigma$ and $\Sigma_m$ \emph{can} have a one dimensional null space that corresponds to the zero mean constraint on the potentials; as will be evident in the next Section, this depends on boundary conditions.

\section{Decomposition}\label{sec:decomposition}

As was evident from the discussion thus far, the decomposition that we espouse relies on imposing solenoidal nature of a component. To do so, we have specified $\Sigma$ and $\Sigma_m$. Mapping to approaches used in the literature is apparent; $\Sigma$ and $\Sigma_m$ are associated with tree's formed either out of edges or faces respectively. The complement of these are the co-trees (of edges and faces). While we shall return to this concept later, the approach we take is more generalizable along the lines of Ref. \cite{andriulli2012well}. This follows the observation that in a Helmholtz decomposition in multiply connected domains, both the irrotational component and the harmonic component are divergence free, and it is \emph{not} important to distinguish between the two. To this end, we introduce projector that effect this decomposition.

\subsection{Projector Matrices}

Development of projectors, given \eqref{eq:efld_decomp} and \eqref{eq:bfld_decomp}, is relatively straightforward. In what follows, we will illustrate this for the electric field. As alluded to earlier, $\Sigma^T \Lambda = 0$ implies $\Sigma^T \vb{E}_s (n\Delta_t, \vb{r})=0$. It follows from  \eqref{eq:efld_decomp_flux} that  
\begin{equation}
  \bar{E}^n_{ns} = \left ( \Sigma^T \Sigma\right )^\dagger\Sigma^T \bar{D}^n  
\end{equation}
where the $\dagger$ denotes a Morse-Penrose psuedo inverse. We note though that abstraction hides insight; in this case, $\Sigma^T \bar{D}^n$ is  related to the Gauss' law for electric fields and can be used to solve for $\bar{E}^n_{ns}$ directly, a tactic we will flesh out in detail in next subsection. Completing the train of thought, one can now define projectors
\begin{subequations}
\begin{align}
    [\vb{P}]_e^{\Sigma} & = \Sigma(\Sigma^T \Sigma)^\dagger \Sigma^{T} \label{eq:proj_es}\\
    [\vb{P}]^{\Lambda}_e & = \mathcal{I} - [\vb{P}]_e^{\Sigma} \label{eq:proj_el}
\end{align}
\end{subequations}
Using these projectors, it follows that \eqref{eq:efld_decomp_flux} can be rewritten as 
\begin{equation}
\label{eq:efld_projected}
    \bar{D}^n = \Sigma \bar{E}^n_{ns} + \left [ \vb{P}\right ] ^\Lambda_e \bar{D}^n
\end{equation}
If, $\bar{E}^n_{ns} $ has already been computed, what remains is to compute the rest of the contributions. Similar approach is possible for the magnetic flux density. Here, the projectors take the form 
\begin{equation}
\label{eq:bfld_projector}
\left [ \vb{P}\right ]^\Lambda_b = \mathbf{I} - \Sigma_m\left ( \Sigma_m^T \Sigma_m \right )^\dagger \Sigma_m^T 
\end{equation}
and the magnetic flux density can be written as 
\begin{equation}
\label{eq:bfld_projected}
    \bar{B}^n = \Sigma_m \bar{B}^n_{ns} + \left [ \vb{P}\right ]^\Lambda_b \bar{B}^n
\end{equation}
It follows that the magnetic Gauss' law is trivially satisfied provided we choose  $\bar{B}^n_{ns} = 0$. As an aside, the motivation to present ideas in terms of projectors is to lay the foundation for extending these to both higher order and make the decomposition genus free. This extension will be the thrust of another forthcoming paper. The main goal of this paper, is to fully develop and demonstrate these ideas for 0$^{th}$ order Whitney basis and simply connected structures.   

\subsection{System of Equations}

\begin{figure}
    \centering
    \includegraphics[scale=0.6]{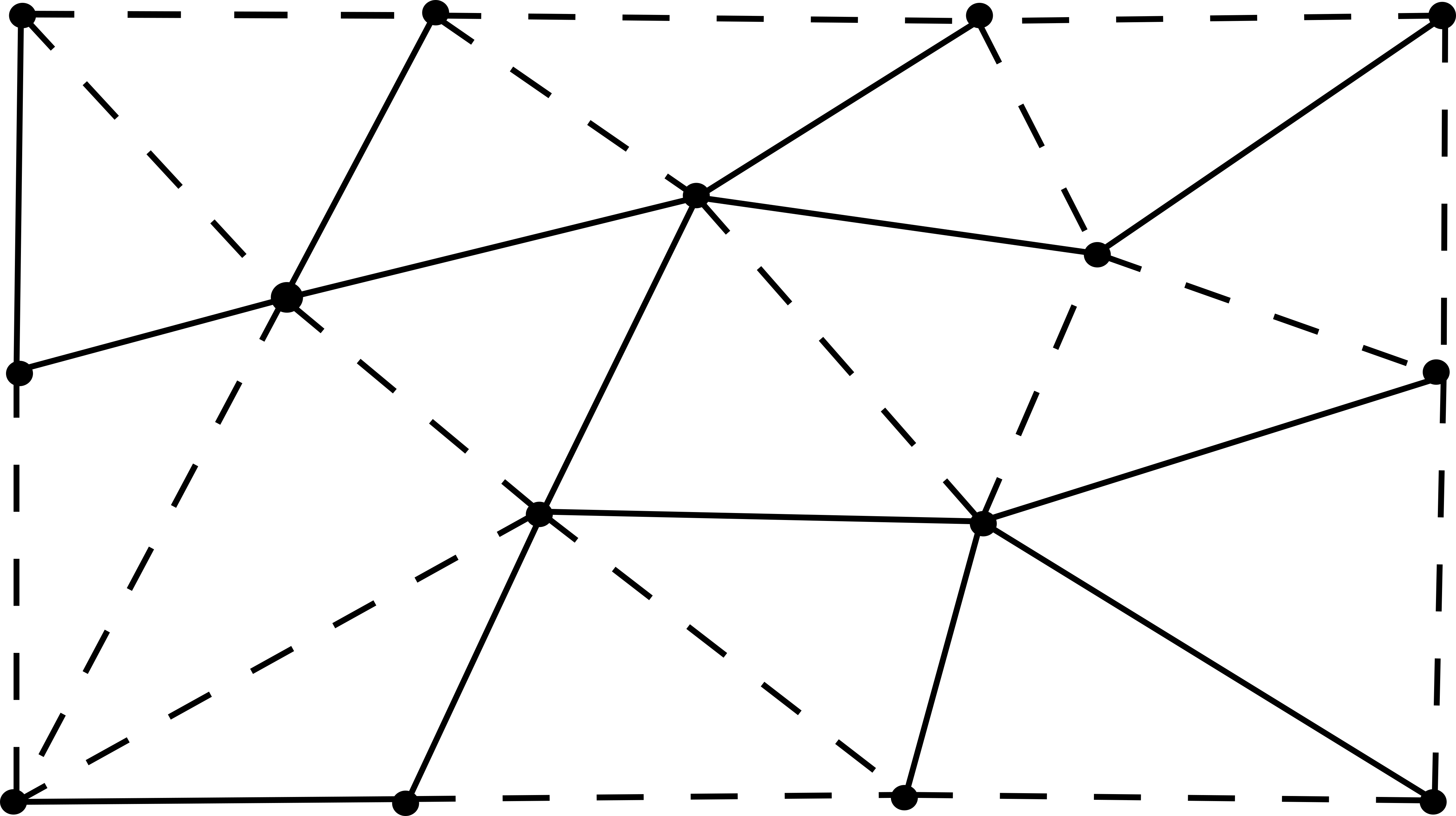}
    \caption{Tree (solid) and cotree (dashed) edges for a 2D geometry.}
    \label{fig:cotree_diagram}
\end{figure}

\begin{figure}
    \centering
    \includegraphics[scale=0.6]{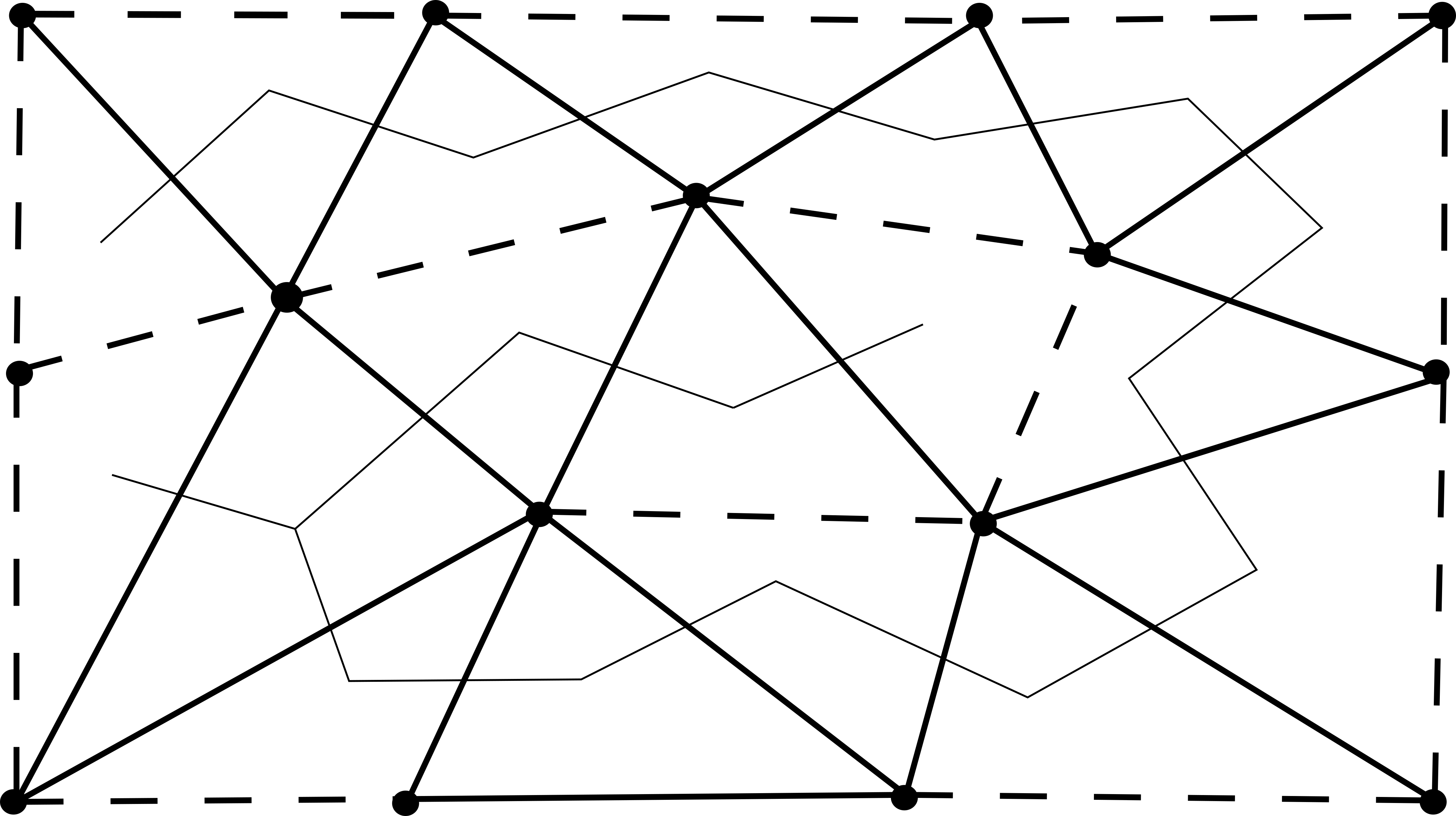}
    \caption{Tree (solid) and cotree (dashed) faces for a simplified 2D tree structure.}
    \label{fig:face_cotree_diagram}
\end{figure}

 Identification of both the non-solenoidal and the solenoidal component has been studied for (non)simply connected structures for a long time \cite{}. In this paper, this forms our starting point. Given any network, one can create a minimum spanning tree as shown in Fig. \ref{fig:cotree_diagram}. From this minimum spanning tree, one can identify both the tree and co-tree edges; specifically, we note that $\Sigma$ maps $\mathcal{N} \longrightarrow \mathcal{E}$ and $\Lambda$ maps $\mathcal{E}_{c} \longrightarrow \mathcal{E}$ where $\mathcal{E}_c \subset \mathcal{E}$. In a similar manner, one can construct a tree (and co-tree) of faces; this is illustrated in two-dimensions in  Fig. \ref{fig:face_cotree_diagram}. Here, $\Sigma_m$ is a map from $\mathcal{T} \longrightarrow \mathcal{F}$, and $\Lambda_m$ is a map from $\mathcal{F}_c \longrightarrow \mathcal{F}$ where $\mathcal{F}_c \subset \mathcal{F}$. Here, and henceforth, the subscript ``c'' will denote co-tree quantities. It follows that one can define matrices
 \begin{equation}\label{eq:cotree_mat_e}
    [\vb{C}^e_c]_{i,j}  = \left \{ 
    \begin{array}{ll}
        1, & i \in \mathcal{E}, j \in \mathcal{E}_c\\
        0, & \text{otherwise} 
  \end{array}  \right . 
\end{equation}
Likewise,
\begin{equation}\label{eq:cotree_mat_b}
    [\vb{C}^b_c]_{i,j}  = \left \{ 
    \begin{array}{ll}
        1, & i \in \mathcal{F}, j \in \mathcal{F}_c\\
        0, & \text{otherwise} 
  \end{array}  \right . 
\end{equation}
A word of caution; if Dirichlet boundary conditions are applied the tree must be grounded to the PEC surface as in Fig. \ref{fig:cotree_diagram}. 
We need one more piece of the puzzle prior to setting up our solution system. As we had noted earlier, the both $\Sigma$ and $\Sigma_m$ have a 1-D null space that arises \textcolor{black}{as a result of boundary condition that need to be imposed; this can be related to the appropriate Whitney spaces being closed as well. For $\bar{E}^n_{ns}$ it manifest itself as zero mean constraint when Neumann boundary conditions are used and for $\bar{B}_c^n$ when Dirichlet boundary conditions as used.} To effect these, we introduce a zero mean constraint matrices $[\vb{C}_z^q]$
\begin{align}\label{eq:zero_mean_constraint}
    [\vb{C}_z^q] = \mqty[1 & 0 & \cdots \\ 0&\ddots&\\ -\frac{\alpha^q_1}{\alpha^q_n}&\cdots&-\frac{\alpha^q_{n-1}}{\alpha^q_n} ].
\end{align}
\begin{align}
    \alpha^k_i = \int_{\Omega} W^{k}_i (\vb{r}) d\vb{r}
\end{align}
for \textcolor{black}{ $q=\left \{e,b \right \}$ and $k = 0$ when $q = e$ and $k = 3$ when $q = b$.} Note, these matrices become identity if the constraint is not necessary; likewise these need be modified when the net charge is not zero. Using this framework, we now examine projectors and solution to our field equations. We note the following: $[\grad]^T\Lambda = 0$ and $[\div] \Lambda_m = 0$ and more to the point, $[\grad]^T \bar{D}^n$ and $[\div] \bar{B}^n$ are the discrete analogs of the continuous Gauss' laws ($[\grad]^T$ is a discrete divergence $[\widetilde{\div}]$ in the dual grid). This fact allows use to redefine projectors; viz, with a slight abuse of notation
\begin{subequations}
\begin{align}
    [ \vb{P}]^\Sigma_e = & \Sigma [\vb{C}_z^e] \left ( [\vb{C}_z^e]^T [\grad]^T \Sigma [\vb{C}_z^e] \right )^{-1} [\vb{C}^e_z ]^T[\grad]^T \label{eq:efld_mod_proj}\\
    [ \vb{P}]^\Sigma_b = & \Sigma_m [\vb{C}_z^b] \left ( [\vb{C}_z^b]^T [\div] \Sigma_m [\vb{C}_z^b] \right )^{-1} [\vb{C}^b_z ]^T[\div] \label{eq:efld_mod_proj}
\end{align}
\end{subequations}
Using these in \eqref{eq:proj_el} and  \eqref{eq:bfld_projector} enables us to derive $[\vb{P}]^\Lambda_e$ and $[\vb{P}]^\Lambda_b$ and then using \eqref{eq:efld_projected} and 
\eqref{eq:bfld_projected} we have a complete description of the electric and magnetic fields/flux densities. 
We  now have the necessary quantities to prescribe a PIC scheme to consistently evaluate $\bar{E}^n_{ns}$, $\bar{E}_s^n$ and $\bar{B}_s^n$ given $\bar{J}^n$. We will, \emph{a-priori}, set $\bar{B}^n_{ns} = 0$. Just to reiterate, there is a slight abuse of notation. Quantities are assumed to evaluated at $n\Delta_t$. We have \emph{not} specified temporal basis sets used for discretization or how functions are measure in time. We will leave that to later in this section when it is more appropriate. 

\subsubsection{Solution for non solenoidal component  $\vb{E}_{ns}^n$}

The starting point of the solution is the discrete Ampere's law at time point $n\Delta_t$ which reads $\partial_t  \bar{D}^n - [\curl]^T[\star_{\mu^{-1}}] \bar{B}^n = - \bar{J}^n$. Using \eqref{eq:efld_projected}. the constraint $[\vb{C}]_z^e$, operating both sides by $[\grad]^T$, and using the identities prescribed earlier ($[\grad]^T[\vb{P}]^\Lambda_e = 0 = [\grad]^T[\curl]^T$) results in
\begin{equation}
\begin{split}
        [\grad]^T\Sigma [\vb{C}_z^e] \partial_t \bar{E}^n_{ns} & = -  [\grad]^T\bar{J}^n \\
        [\grad]^T [\star_\varepsilon][\grad][\vb{C}_z^e] \bar{E}^n_{ns} & = - [\grad]^T\int_0^{n\Delta_t} \bar{J} (\tau)d\tau = \bar{\rho}^n
\end{split}
\label{eq:discreteGauss}
\end{equation}
It is apparent that \eqref{eq:discreteGauss} is the discrete Gauss' law for the electric fields. The left hand side of this equation is the discrete Laplacian, and coefficients $\bar{E}_{ns}^n$ are the appropriate values of the potential due to the charge density on the right hand side at that instance of time. Further note, as shown in \cite{crawford2021rubrics} this equation can be rewritten as, 
\begin{equation}
    \bar{G}^n_i = \int_0^{n\Delta_t} j_i (t) d\tau = \int_{\vb{r}(0)}^{\vb{r}(n\Delta_t)} \langle \vb{W}_i^{(1)} (\tilde{\vb{r}}), \vb{J} (t, \tilde{\vb{r}}) \rangle d\tilde{\vb{r}} 
\end{equation}
It follows that, $\bar{\rho}^n = -[\grad]^T \bar{G}^n$.  From the equation above, it follows that the integration over time is subsumed in the path integral used to evaluate the charge density\cite{crawford2021rubrics,2021arXiv210206248O}. As a result, one obtains the solution to the Laplacian directly. Two points that should be noted: (a) the equation of continuity is analytically satisfied, and (b) the electric flux density is obtained by explicitly solving Gauss' law. Having obtained $\bar{E}_{ns}^n$, obtaining the rest is fairly straightforward for \emph{both} MFEM and WE-FEM. 

\subsubsection{Solution to the solenoidal components}

Having obtained the solution to the non-solenoidal components, the solution to the rest is relatively straightforward. We begin with rewriting Maxwell's equations as 
\begin{subequations}
\label{eq:reduceMEs}
\begin{align}
    [\vb{Z}]_{11}\partial_t  \bar{B}^n_{s} + [\vb{Z}]_{12} \bar{E}^n_{s} & = -[\vb{Z}]_{13} \bar{E}^n_{ns} \label{eq:reduced_faradaysLaw}\\ 
    [\vb{Z}]_{21} \partial_t \bar{E}^n_{s} - [\vb{Z}]_{22} \bar{B}^n_s & = - \partial_t \bar{G}^n - [\vb{Z}]_{23} \partial_t \bar{E}^n_{ns} \label{eq:reduced_ampereLaw}
\end{align}
\end{subequations}
where
\begin{subequations}
\begin{align}
    [\vb{Z}]_{11} = &  [\vb{C}_c^b]^T[\vb{P}]_b^\Lambda [\vb{C}_c^b] \\
    [\vb{Z}]_{12} = & [\vb{C}_c^b]^T [\curl] [\star_\varepsilon]^{-1}[\vb{P}]_e^\Lambda [\star_\varepsilon] [\vb{C}_c^e] \\
    [\vb{Z}]_{13} = & [\vb{C}_c^b]^T [\curl] [\star_\varepsilon]^{-1}\Sigma [\vb{C}_z^e]\\
    [\vb{Z}]_{21}  = & [\vb{C}_c^e]^{T}[\vb{P}]_e^\Lambda[\star_\varepsilon][\vb{C}_c^e] \\
    [\vb{Z}]_{22} = & [\vb{C}_c^e]^{T} [\curl]^T [\star_{\mu^{-1}}][\vb{P}]_b^\Lambda [\vb{C}_c^b] \\
    [\vb{Z}]_{23} = & [\vb{C}_c^e]^{T} \Sigma  [\vb{C}_z^e]
\end{align}
\end{subequations}
Since $\Sigma =[\star_\varepsilon] [\grad] $, it is trivial to show that $[\vb{Z}]_{13} = 0 $. The above is a re-writing of Maxwell's equation on the tetrahedral grid. Using the above, the wave equation can be trivially derived using \eqref{eq:reduced_faradaysLaw} in the time derivative of \eqref{eq:reduced_ampereLaw}. Specifically, using the above notation, these can be written as
\begin{equation}
    \label{eq:reducedWaveEqn}
    \begin{split}
        [\vb{Z}]_{21}\partial_t^2\bar{E}^n_s & + [\vb{Z}]_{22}[\vb{Z}]^{-1}_{11} [\vb{Z}]_{12} \bar{E}^n_s  = - \partial_t^2 \bar{G}^n \\
        & - [\vb{Z}]_{23}\partial_t \bar{E}^n_{ns} 
    \end{split}
\end{equation}
These equations form the backbone of both our Maxwell and wave equation solvers. As presented we have not presented any specific scheme to evolve both the field equation or the particles. In the results presented, we have used Newmark-$\beta$ to evolve the fields and a fourth order Adams method to map the trajectory and velocity of the particles. These are developed detail in Ref. \cite{2021arXiv210206248O} and are not repeated here.

\section{Null spaces, gauging and other remarks}\label{sec:nullSpaces}

The methodology presented in this paper has several advantages over conventional FEM-PIC; one of the fundamental challenges with time domain FEM is the presence of a null-space. It has been shown the leap-frog mixed FEM (MFEM) does not suffer from a null space when its initial conditions are specified. But the downsides of the leapfrog MFEM is that it is conditionally stable with a mesh-dependent time step size. MFEM with Newmark overcomes this fundamental bottleneck; time step size can be arbitrary large (\emph{albeit} it needs to be chosen to capture the physics). Unfortunately, there is a null space that  behaves as $\grad \phi (\vb{r})$ where $\phi (\vb{r})$ is some scalar function. The level of excitation of this null space depends on methods used to invert the matrices in the time marching system. In both the quasi-static limit and high tolerance of the iterative solver, this is small. Unlike MFEM, the null space for the wave equation FEM (WE-FEM) grows like $t\grad (\vb{r})$. The effect of this on charge conservation is evident in Ref. \cite{2021arXiv210206248O}. 

What is abundantly clear from the above, is that null space of both MFEM and WE-FEM play a significant role on the spurious numerical artifacts that manifest themselves in conservation laws. From this perspective, we note the following: 
\begin{enumerate}
    \item The solutions for $\bar{E}^n_{ns}$ does not depend on a time integrator on the left hand side. This is effected on the RHS when integrating over the path.
    \item The resulting equation for $\bar{E}^n_{ns}$ is exactly the discrete Gauss' law with the equation of continuity used to create the RHS.
    \item The null space of these equation are harmonic solutions that are not excited provided boundary conditions are properly imposed.
    \item The solution to both $\bar{E}_s^n$ and $\bar{B}_s^n$ have null spaces that are of the form $\grad \phi (\vb{r})$.
    \item Despite the presence of the null space, \emph{by design}, $[\div] \bar{B}^n = 0$ and the divergence of the solenoidal component to the electric flux density $[\grad]^T [\vb{P}]^\Lambda_e [\star_\varepsilon] [\vb{C}]_c^e \bar{E}_{s}^n = 0$. 
    \item As a result, conservation laws  are satisfied and not corrupted by null spaces. 
    \item Parenthetically, we note that the number of degrees of freedom of $\bar{E}_s^n$ and $\bar{B}_{s}^n$ is identical for simply connected domains (via Euler's relations). These may be useful for some symplectic PIC schemes.    
\end{enumerate}

Next, we note that the above scheme bears a strong resemblance to the well known $\vb{A}-\Phi$  formulation. But we note some not-so-subtle differences. Note standard decomposition relies on the electric field. We note that the decomposition in this paper relies on the electric flux density, and is \emph{not} a complete Helmholtz decomposition (which is not possible in a discrete setting). Decomposition of the flux density induces a mesh based metric in mapping between the primal and dual grids (can thought of simple mesh based anisotropy). Using this is imperative in solving Maxwell's equations. Gauge condition  (Coulomb) are, however, identical; likewise, final system of equations bears a strong similarity to those obtained in an $\vb{A} -\Phi$ formulation. By virtue of our accuracy in maintaining the continuity equation, the use of the Coulomb gauge does not adversely impact our numerical description of the space-charge and space-current sources\cite{lau1992review}.

\FloatBarrier
\section{Results}\label{sec:results}

In what follows, we present a sequence of results to verify the methodology presented earlier. We will show that (a) the solution to \eqref{eq:discreteGauss} does not have a null space provided one imposes a zero mean constraint and the solution to  \eqref{eq:reduceMEs} has a null space corresponding to a gradient of a time invariant scalar potential. In the course of this test, we will also show that the solution to \eqref{eq:reduceMEs} is unconditionally stable. Next, we will test the proposed methodology against two test problems and validate the behaviour of the solution.

\subsection{Field Solver Validation}

\subsubsection{Eigenvalues}
\begin{figure}
    \centering
    \includegraphics[scale=0.5]{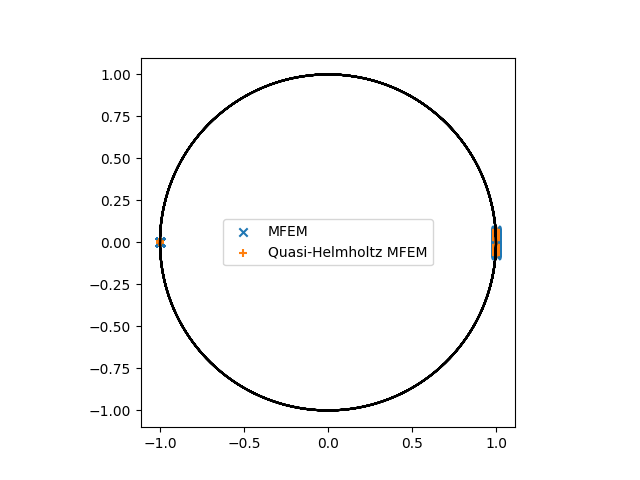}
    \caption{Eigenvalues Newmark-beta time stepping of MFEM and solenoidal-MFEM system for PEC box.}
    \label{fig:eigenvalues}
\end{figure}

The temporal solution of \eqref{eq:reduceMEs}  utilizes Newmark-beta as a time stepping method. To verify that the resulting system is unconditionally stable and examine its null space, we find the eigenvalues of the discrete system as described in Ref.\cite{crawford2020unconditionally}. Our discrete system is obtained from discretizing a cube of side length 1$m$ using tetrahedra with average edge length of 0.214$m$. Dirichlet boundary condition ($\hat{n} \cross \vb{E}(t, \vb{r}) = 0)$ is imposed on the walls. The eigenvalues for reduced system in \eqref{eq:reduceMEs} is compared against those obtained for a full MFEM solve. As is evident from  Fig. \ref{fig:eigenvalues} eigenvalues for both systems lie on the unit circle making the system energy conserving and unconditionally stable. Furthermore, the presence of eigenvalues at zero indicate a null space that behaves as a gradient of potential, a feature that both methods share. 
\FloatBarrier

\subsubsection{Neumann Boundary Condition}

Next, we examine the behavior of the system under Neumann boundary conditions. Recall that one has impose a zero mean constraint on $E^n_{ns}$. This test validates proper imposition of this condition as well as the solver as a whole. As before, our domain is a cube with side length 1$m$. 
The mesh is discretized with an average edge length of 0.429m yielding 133 tetrahedral elements. Neumann boundary conditions are equivalent to a plane wave polarized along $-\hat{z}$ and propagating along $\hat{x}$ direction, with temporal variation being the derivative of modulated Gaussian with center frequency of 10 MHz and bandwidth of 5 Mhz. The explicit formulae is fairly standard and can be obtained from say\cite{crawford2020unconditionally}.  

\begin{figure}
    \centering
    \includegraphics[scale=0.5]{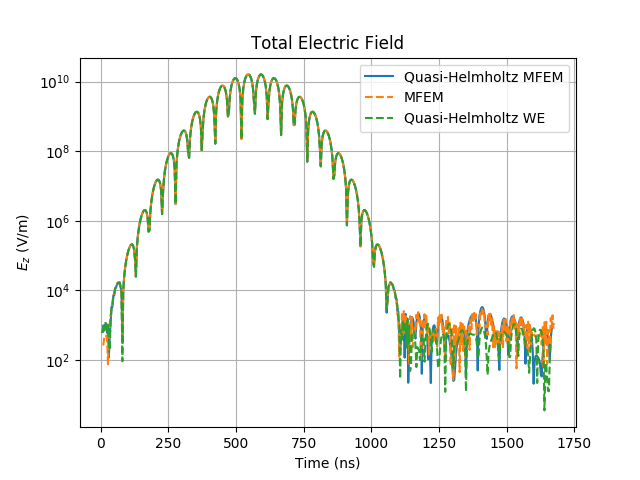}
    \caption{Total electric field for plane wave through a box with Neumann boundary conditions.}
    \label{fig:plane_wave_total_field}
\end{figure}

\begin{figure}
    \centering
    \includegraphics[scale=0.5]{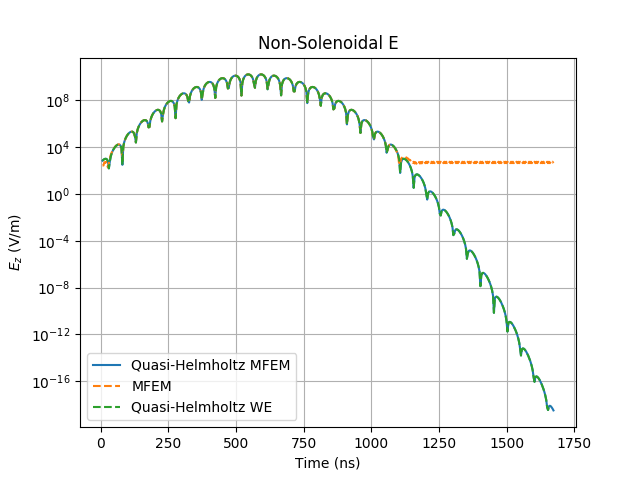}
    \caption{Non-solenoidal electric field for plane wave through a box with Neumann boundary conditions.}
    \label{fig:plane_wave_non-solenoidal}
\end{figure}

\begin{figure}
    \centering
    \includegraphics[scale=0.5]{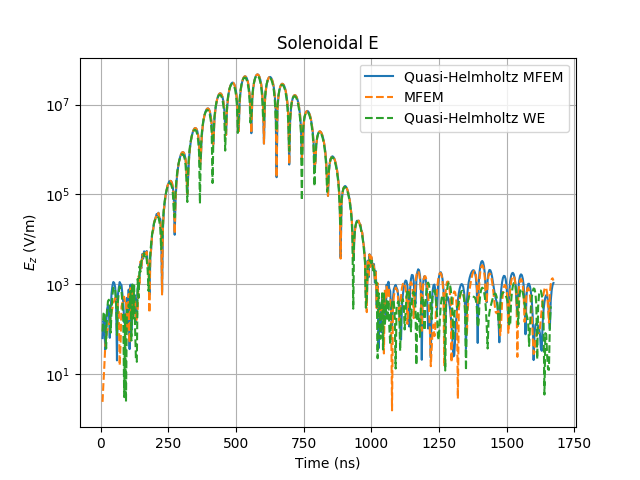}
    \caption{Solenoidal electric field for plane wave through a box with Neumann boundary conditions.}
    \label{fig:plane_wave-solenoidal}
\end{figure}

We note the following. Given results from a standard MFEM solve, we can use \eqref{eq:efld_decomp} to partition into solenoidal and non-solenoidal components. That is, given coefficients $\bar{E}^n$ we can obtain $\bar{E}^n_{ns}$ and $\bar{E}^n_{s}$. Alternatively, we can solve \eqref{eq:discreteGauss} for $\bar{E}^n_{ns}$ and either \eqref{eq:reduceMEs} or \eqref{eq:reducedWaveEqn} for $\bar{E}^n_s$. So we compare can both the component wise solution from the reduced system as well as the full solution. First at a random unknown in computational domain, we compared the total electric field from both the quasi-Helmholtz decomposition and MFEM system in Fig. \ref{fig:plane_wave_total_field}. 
Next, the solutions from the standard MFEM system is decomposed solenoidal and non-solenoidal components and compared with the solendoidal and non-solenodial components of the quasi-Helmholtz system. In Fig \ref{fig:plane_wave_non-solenoidal} and \ref{fig:plane_wave-solenoidal} the non-solenoidal components and solendodial components are compared. 

A couple of points to note: (a) from Fig. \ref{fig:plane_wave_non-solenoidal} the solution of classical MFEM has a null space (as expected) whereas that of \eqref{eq:discreteGauss} does not. (b) The solution of solenoidal components of all three have a null space, in that they do not smoothly go to zero once the excitation vanishes. (c) The agreement between all three is excellent (approximately single precision) in the regime of interest despite all three being different systems. 

\subsection{Particle Beam}

Having validated the field solution, we turn our attention to inclusion of particles in the system. We leverage the test bank developed in Ref. \cite{2021arXiv210206248O} for examples. In the first test case electron macro particles are injected into a cylindrical cavity. 
The electrons repel away from each other and expand. 
The profile of the trajectories can be compared with other verified methods as well as quasi-analytic solutions\cite{o2021set}. Note, in this case  $\bar{E}_{s}^n$ will correspond to cavity modes and $\bar{E}_{ns}^n$ that will be the dominant component. Details of the experiment are provided in \ref{tb:beam}. 

\begin{figure}
    \centering
    \includegraphics[scale=0.5]{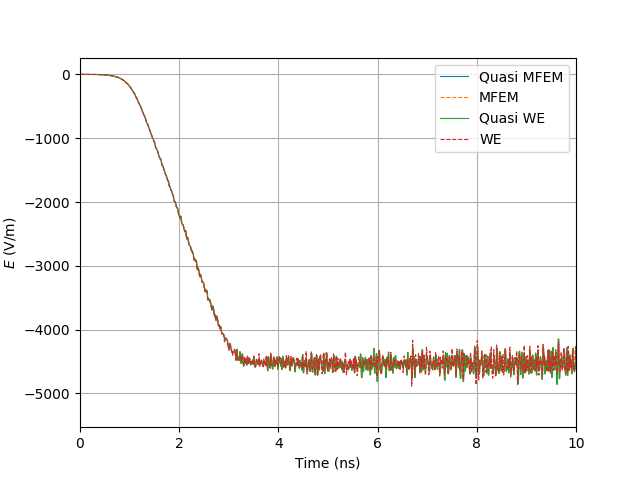}
    \caption{Comparison of radial electric field $\bar{E}$ between MFEM, qausi-MFEM, wave equation, and quasi wave equation.}
    \label{fig:cotree_vs_newmark}
\end{figure}

In Fig. (\ref{fig:cotree_vs_newmark}), the electric field in the radial direction half way down the tube 16 mm in the radial direction is compared with the same run using a Newmark-beta formulation with and without the quasi-Helmholtz decomposition (data for without from\cite{2021arXiv210206248O}). The agreement between all three sets of data is excellent.  

Next, we can examine each of the components. To wit, as before, we decompose $\bar{E}^n$ from a full MFEM solve into $\bar{E}^n_{ns}$ and $\bar{E}^n_s$ and compare this data against that obtained from the reduced systems presented here (both \eqref{eq:reduceMEs} and \eqref{eq:reducedWaveEqn}). 

\begin{figure}
    \centering
    \includegraphics[scale=0.5]{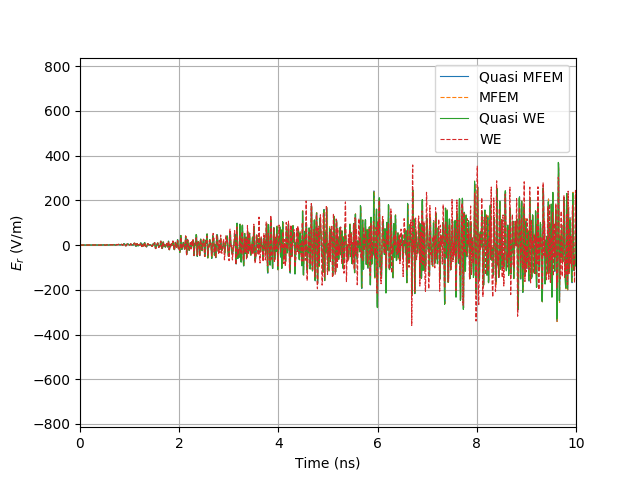}
    \caption{Comparison of radial rotational electric field $\bar{E}_{s}$ between MFEM, qausi-MFEM, wave equation, and quasi wave equation.}
    \label{fig:cotree_vs_newmark_ec}
\end{figure}

\begin{figure}
    \centering
    \includegraphics[scale=0.5]{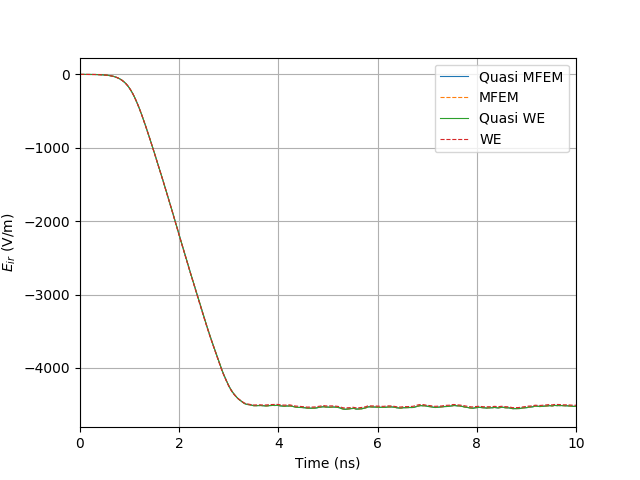}
    \caption{Comparison of radial irrotational electric field $\bar{E}_{ir}$ between MFEM, qausi-MFEM, wave equation, and quasi wave equation.}
    \label{fig:cotree_vs_newmark_es}
\end{figure}

\begin{figure}
    \centering
    \includegraphics[scale=0.5]{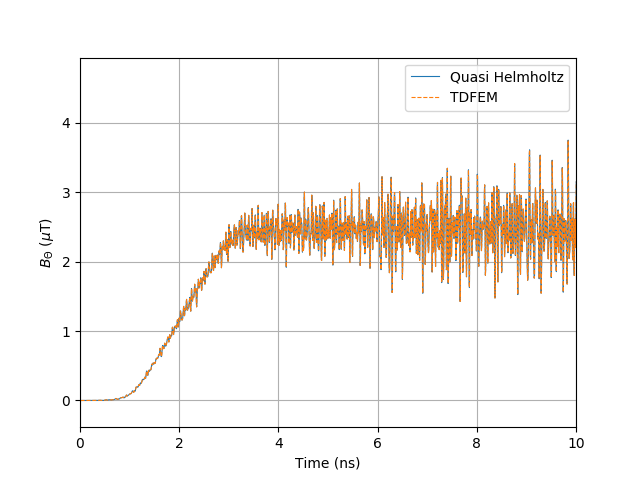}
    \caption{Comparison of radial cotree magnetic field between cotree solve and newmark solve.}
    \label{fig:cotree_vs_newmark_b}
\end{figure}

First, as is evident from Fig. \ref{fig:cotree_vs_newmark_ec} compares the $\bar{E}_{s}^n$ components, they contain no DC components and contain all the cavity modes. The $\bar{E}^n_{ns}$ component can be seen in \ref{fig:cotree_vs_newmark_es} is the DC component shown in the total field Fig. \ref{fig:cotree_vs_newmark}. Next, the magnetic field shown in Fig. \ref{fig:cotree_vs_newmark_b} shows excellent agreement between the magnetic field obtained from solution with and without quasi-Helmholtz decomposition. 

\begin{figure}
    \centering
    \includegraphics[scale=0.5]{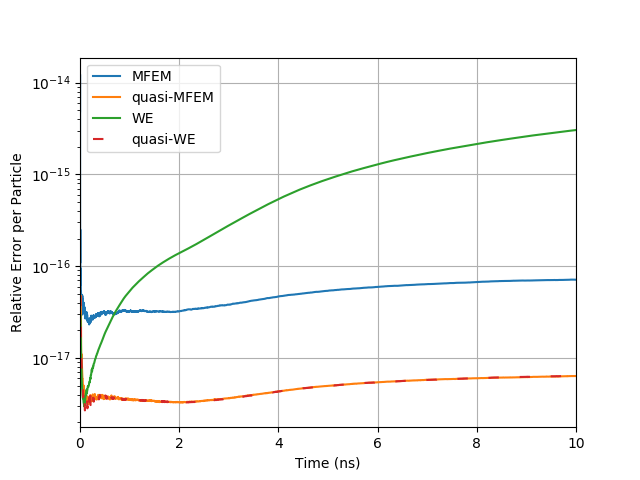}
    \caption{Comparison between regular mixed finite element, wave equation, and quasi-Helmholtz wave equation error in the discrete continuity equation. The growing in time DC null space in the wave equation is remedied by the quasi-Helmholtz decomposition.}
    \label{fig:DGL}
\end{figure}

The removal of the growing in time $t\grad\Phi(\vb{r})$ null space for the wave equation is of considerable interest. In Fig. \ref{fig:DGL} we compare the MFEM and wave equation error in discrete Gauss's law \emph{without} quasi-Helmholtz projectors. As evident, the error grows with time for WE-FEM and thresholds at some value for MFEM. With the quasi-Helmholtz decomposition, the story is very different. As is to be expected, the error for MFEM and WE-FEM is identical and about a order below the results obtained earlier. 

\begin{table}[ht!]
\centering
    \caption{Expanding Particle Beam Parameters}
    \begin{tabular}{c|c}
         \textit{Parameter} & \textit{Value}  \\
         \hline
         Cavity Radius & 20 mm \\
         Cavity Length & 100 mm\\
         Boundary Conditions & PEC \\
         $v_p$ & $5\cdot10^7$ m/s \\ 
         $v_p/c$ & 0.16678 \\ 
         beam radius $r_b$ & 8.00 mm \\
         Number particles per time step & 10 \\
         species & electrons \\
         Turn on time & 2 ns \\
         beam current  & 0.25 A \\
         macro-particle size & 52012.58 \\
         min edge length & 1.529 mm \\
         max edge length & 6.872 mm \\
    \end{tabular} \label{tb:beam}
\end{table}

\subsection{Adiabatic Expansion of Plasma}

\begin{figure}
    \centering
    \includegraphics[scale=0.5]{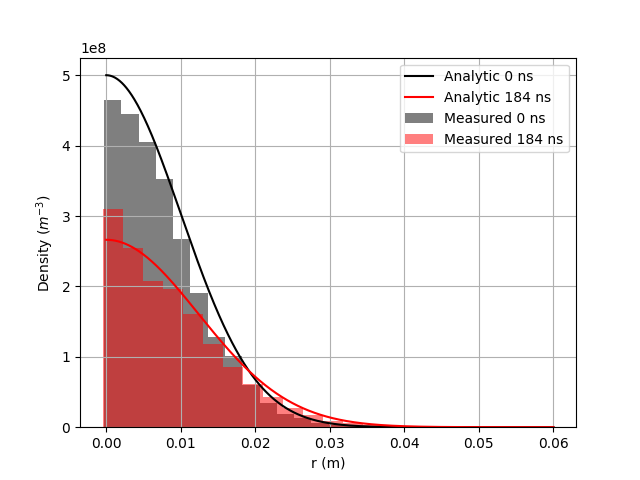}
    \caption{Quasi-Helmholtz decomposition adiabatic expansion using quasi-MFEM.}
    \label{fig:plasma_ball_quasi}
\end{figure}

The final validation case is the adiabatic expansion of a plasma ball. This test case has analytic solutions and allows for good comparison and validation\cite{kovalev2003analytic}. In this example,  electrons and ions are placed in the center of a mesh with a Gaussian radial distribution. 
Ions are given a low temperature 1k and the electrons 100k. The system is initially charge neutral.  The electrons initially expand outward creating an electric field that pulls the ions outward. The density over time has analytic solutions and is presented along with the measured density from the simulation in Fig. \ref{fig:plasma_ball_quasi}. 
Details of this test example can be found in \cite{o2021set} and not repeated here. As is evident, the proposed solution agrees well with analytic data.

\section{Conclusion}

In this paper, we have taken a step forward in building a FEM-PIC scheme that uses implicit field solvers and is robust to corruption due to null spaces. Our approach has been to define quasi-Helmholtz projectors, and in doing so, we show (a) satisfaction of conservation laws as well as (b) correctness against data obtained earlier. An important item to note is that the dimension of our system, in terms of number of degrees freedom, is slightly smaller than earlier. Of course, this is a foundational paper in that we have not addressed other pressing issues such as complexity, extension to higher order basis, multiply-connected domain, sympletic methods, and so on. Work on some of these topics is underway and will be presented elsewhere. 

\begin{acknowledgments}

This work was supported by SMART Scholarship program. We thank the MSU Foundation for support through the Strategic Partnership Grant during early portion of this work. This work was also supported by the Department of Energy Computational Science Graduate Fellowship under grant DE-FG02-97ER25308 and financial support from NSF via CMMI-1725278. The authors would also like to thank the HPCC Facility,
Michigan State University, East Lansing, MI, USA.
\end{acknowledgments}

\section*{Data Availability \label{sec:data}}
The data that support the findings of this study are
available from the corresponding author upon reasonable
request.

\bibliography{helmholtz}

\end{document}